\documentclass[amsmath,amssymb,superscriptaddress,twocolumn,reprint]{revtex4}
\usepackage[pdftex]{graphicx}
\usepackage{upgreek}
\begin{document}
\title{High sensitivity optomechanical reference accelerometer over 10\,kHz}
\author{Felipe Guzm\'an Cervantes$^{\dagger}$}\thanks{These authors contributed equally to this work.}
\author{Lee Kumanchik}\thanks{These authors contributed equally to this work.}
\affiliation{National Institute of Standards and Technology, Gaithersburg, MD\,20899, USA.}
\affiliation{Joint Quantum Institute, University of Maryland, College Park,
MD\,20742, USA.}
\author{Jon Pratt}
\affiliation{National Institute of Standards and Technology, Gaithersburg, MD\,20899, USA.}
\author{Jacob\,M. Taylor}\email[Corresponding authors:\\]{felipe.guzman@nist.gov, jacob.taylor@nist.gov}
\affiliation{National Institute of Standards and Technology, Gaithersburg, MD\,20899, USA.}
\affiliation{Joint Quantum Institute, University of Maryland, College Park,
MD\,20742, USA.}
%
\begin{abstract}
We present an optically-detected mechanical accelerometer that achieves a sensitivity of 100\,$\mathrm{n}g_n/\sqrt{\mathrm{Hz}}$ over a bandwidth of 
10\,kHz and is traceable. We have incorporated a Fabry-P\'erot fiber-optic micro-cavity that is currently capable of measuring the test-mass displacement with sensitivities of $200\,\mathrm{am}/\sqrt{\mathrm{Hz}}$, and whose length determination enables traceability to the International System of Units (SI). The compact size and high $mQ$-product achieved combined with the high sensitivity and simplicity of the implemented 
optical detection scheme highlight our device and this category 
of accelerometers, outlining a path for high sensitivity reference acceleration measurements and observations in seismology and gravimetry.
\end{abstract}
\maketitle
%
The reflection spectrum of an optical cavity is exquisitely sensitive to motion-driven frequency changes  which, when combined with mechanical 
oscillators, can yield accelerometers of outstanding observation resolution and 
bandwidth. Typically, acceleration is obtained from a direct displacement measurement, and it is in the sensitivity of this observation $x_\mathrm{res}$ where the fundamental compromise lies of reaching a particular acceleration resolution $a_\mathrm{res}$ over a certain bandwidth $\omega_o$: $x_\mathrm{res}=a_\mathrm{res}/\omega_o^2$. The higher the 
sensitivity and the larger the bandwidth needed in acceleration sensing, the more 
challenging will be the required displacement resolution. Mechanical properties 
of the oscillator such as the stiffness (mass and resonance frequency 
$\omega_o$) and quality factor $Q$ also play an important role in the sensing 
limit and transduction of displacement measurement to acceleration, as is 
explained below. Particularly, for high performance 
acceleration sensing at room temperature, a combined product of large oscillator 
mass and high quality factor $mQ\geq1\,$kg will be needed to achieve resolutions at sub-$\upmu g_n/\sqrt{\mathrm{Hz}}$ ($1\,g_n=9.80665\,\mathrm{m/s^2}$) over several kHz.

In experimental gravitational physics, remarkably high acceleration resolutions 
at levels of $\mathrm{f}g_n/\sqrt{\mathrm{Hz}}$ have been demonstrated by 
torsion balances over extremely narrow bandwidths below 10\,mHz, which 
corresponds to a displacement resolution of approximately 
10\,$\mathrm{pm}/\sqrt{\mathrm{Hz}}$ \cite{BIBREF1,BIBREF2}. In geodesy and 
geophysics, superconducting gravimeters reach acceleration resolutions of the 
order of $\mathrm{p}g_n/\sqrt{\mathrm{Hz}}$ over bandwidths below 1\,Hz at 
displacement resolutions of tens to hundreds of $\mathrm{pm}/\sqrt{\mathrm{Hz}}$ 
\cite{BIBREF3,BIBREF4}. In contrast, acceleration measurements over wider bandwidths of 
10-30\,kHz have been realized, however, at tens of 
$\upmu g_n/\sqrt{\mathrm{Hz}}$ levels, translating to a sensitivity in 
displacement measurement of a few $\mathrm{fm}/\sqrt{\mathrm{Hz}}$ 
\cite{BIBREF5}. Traditional applications require either high 
acceleration resolution, such as in gravimetry or seismology well below 
100\,Hz, or large bandwidths, as for characterizing fast mechanical dynamics 
and piezo-electric devices, but typically not both.
Yet extremely high resolution maintained over large bandwidths of tens of kHz 
are of interest in applications such as inertial navigation of fast moving 
objects\cite{BIBREF5a}.

Present state-of-the-art accelerometer calibrations employ carefully designed 
and instrumented shaker tables that move with prescribed accelerations that are 
measured using laser interferometry and currently reach levels of 
$10^{-2}$\cite{BIBREF5b} and only at discrete frequency points. The prescribed 
motion realizes an acceleration that is used to calibrate reference 
accelerometers affixed to the tables. The reference accelerometers may then be 
used to calibrate other accelerometers via a back-to-back comparison technique 
executed using shaker tables, each optimized for specific frequency 
ranges\cite{BIBREF5c,BIBREF5d}. Reference accelerometers are mostly used by 
primary calibrations laboratories in a variety of heavy manufacturing 
industries, including aerospace and automotive, to calibrate instruments 
subsequently used in the field.

Here we present an optomechanical reference device which is directly traceable to the  International System of Units (SI) for absolute acceleration sensing that combines both wide bandwidth and exquisite resolution at room temperature, which we have demonstrated to unprecedented sensitivities better than 100\,$\mathrm{n}g_n/\sqrt{\mathrm{Hz}}$ over a 
measurement bandwidth larger than 10\,kHz, between (1.5-12\,kHz). It consists of a 
compact\,($10.6\times15\times2$\,mm), high-$mQ$\,($1$\,kg) fused-silica 
oscillator that utilizes fiber-optic micro-mirror cavities, for traceable 
detection of its test-mass motions\cite{BIBREF6,BIBREF6a}. Since it does not 
require an external shaker to calibrate its sensitivity due to the built-in 
laser interferometer as traceable reference, this device provides substantial 
improvement over conventional systems in accelerometry standards and 
calibrations.

Fundamentally, acceleration resolution is limited by thermal fluctuations of 
the test-mass, which for a simple harmonic oscillator at high temperature is 
given by,\cite{BIBREF7}
\begin{equation}
a_{th}=\sqrt{\frac{4k_BT\omega_o}{mQ}},
\label{Eqn1}
\end{equation}
where $k_B$ is the Boltzmann's constant, $T$ is temperature, $\omega_o$ is the
natural frequency of the harmonic oscillator, $m$ is the oscillator mass, and
$Q$ is the mechanical quality factor. This shows 
that high-resolution (low $a_{th}$) and wide-bandwidth (large $\omega_o$) 
acceleration sensing at room temperature, requires a large $mQ$ product, 
requiring both large mass and high mechanical quality factor. Acceleration and 
displacement are related by the following transfer 
function\cite{BIBREF8,BIBREF9},
\begin{equation}
\frac{X(\omega)}{A(\omega)}=-\frac{1}{\omega_o^2-\omega^2+i\,\frac{\omega_o}{Q}
\omega},
\label{Eqn2}
\end{equation}
where $X(\omega)$ is the relative displacement given an input acceleration
$A(\omega)$ at an angular frequency $\omega$. 

From Equation~\ref{Eqn2}, $100\,\mathrm{n}g_n/\sqrt{\mathrm{Hz}}$-resolution 
over a 10\,kHz-bandwidth requires displacement detection sensitivities at levels 
of $10^{-16}\,\mathrm{m}/\sqrt{\mathrm{Hz}}$. Optical detection schemes, such 
as Fabry-P\'erot interferometry, offer this type of resolution, with the added
benefit of SI wavelength traceability for absolute displacement sensing. Thus, if the test-mass motion is measured in terms of an optical cavity referenced to a physical standard such as an atomic transition, according to Equation~\ref{Eqn2}, the only requirement to convert $X(\omega)$ to acceleration is an accurate enough measurement of $\omega_o$ and $Q$. Current wavelength and frequency calibrations reach relative accuracies of $10^{-11}$ and $10^{-15}$ respectively.

Conventional systems are not able to accurately measure near the mechanical 
resonance due to the complex system identification 
required\cite{BIBREF10,BIBREF11}. Therefore, sensitivity enhancements are 
typically achieved by lowering the sensor's natural frequency and limiting their 
detection bandwidth significantly below the mechanical resonance (typically 
$\approx\omega_o/5$). Due to the accurate and straightforward system identification capabilities presented in Equation~\ref{Eqn2}, our optomechanical reference accelerometer provides traceable measurements at very high sensitivities throughout the entire 
measurement bandwidth about and even beyond the mechanical resonance, where it 
reaches the thermal limit over a bandwidth of approximately 1\,kHz.

\begin{figure}[htp]
\centering
\includegraphics[width=0.365\textwidth, height=0.25\textwidth]{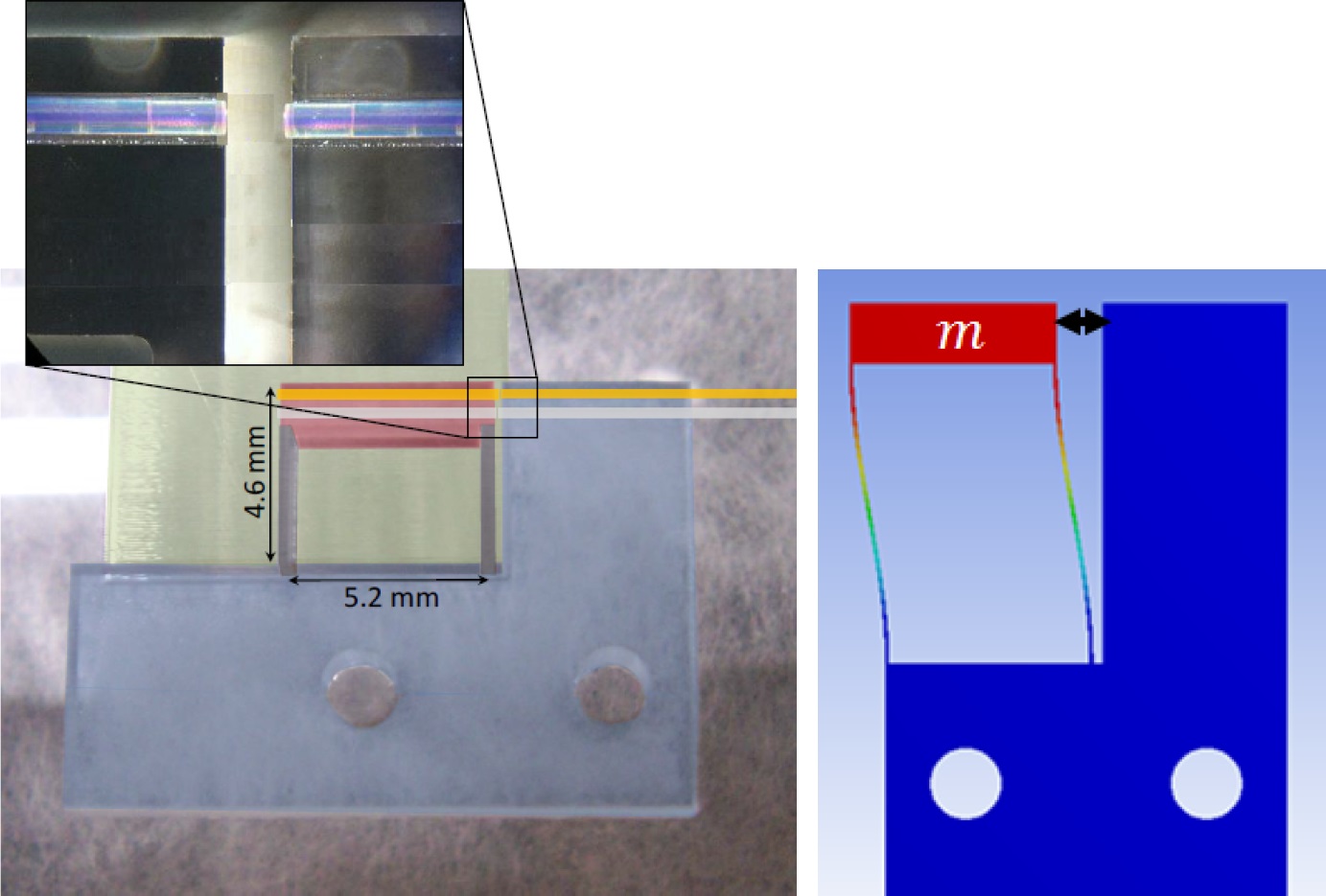}
\caption{\label{figSensor}Photograph (left) and sketch (right) of the monolithic fused-silica mechanical oscillator with the integrated fiber micro-cavity (magnified in the upper left corner).}
\end{figure}

The device we present here combines a 
monolithic fused-silica oscillator and a fiber-optic micro-cavity into an 
absolute reference accelerometer, capable of reaching a resolution below 
100\,$\mathrm{n}g_n/\sqrt{\mathrm{Hz}}$ over 10\,kHz. The direct link between 
the acceleration sensing and the SI-traceable laser wavelength provides 
capabilities for absolute measurements, also around and above the mechanical 
resonance. This is possible because the displacement-to-acceleration transfer 
function (Equation~\ref{Eqn2}) depends only on two parameters, $Q$ and 
$\omega_o$, which are independently measured via a ringdown technique, providing 
a complete system identification.
%
The mechanical fused-silica oscillator is shown in Figure\,\ref{figSensor}.
This material was chosen for its compatibility with fiber optics and its 
inherent low loss characteristics ($fQ>10^{13}$\cite{BIBREF13} or $Q>10^9$ at 
10\,kHz). The mechanical oscillator was laser-assist micro-machined from\cite{BIBREF14} a 2\,mm thick fused-silica wafer to produce a monolithic device. The base of the 
oscillator was frit-bonded to a fused-silica substrate and the moving portion 
was suspended over a laser-machined relief. The mechanical quality factor was 
measured to be $Q=4.2\times10^4$, applying a ringdown technique in vacuum. To this end, we used a piezo-shaker to excite the oscillator at its resonance frequency and measured the exponential decay response. The resonance peak was determined by spectral analysis and then honed in by a high resolution function generator driving the piezo-shaker. After 
discontinuation of shaking, the AC voltage output of the cavity photoreceiver 
was recorded by a digital voltmeter which provided the envelope of the ring down 
directly. The quality factor was determined by an exponential fit to the envelope and compared to the full width at half maximum (FWHM) with good agreement. This technique is 
significantly faster than a high resolution spectrum allowing us to minimize the 
effect of resonance frequency drifts and enabling multiple trials. The moderate 
quality factor measured can be attributed to losses in the frit-bonding 
material. The mass of the oscillator is 25\,mg, 
$\omega_o=2\pi\times10\,710$\,$\mathrm{rad/s}$ and $mQ=1$\,kg, yielding a 
thermal noise floor of $a_{th}\approx3$\,$\mathrm{n}g_n/\sqrt{\mathrm{Hz}}$ at 
room temperature. 

In the optical detection scheme, we utilized a telecom laser at a wavelength of $\lambda=1550$\,nm with moderate 
power of approximately 1\,mW. A plane-concave fiber-based micro-mirror 
Fabry-P\'erot cavity\cite{BIBREF14a,BIBREF15} was built to operate in reflection as the optical sensor. The mechanical oscillator ground platform and the moving mass were equipped with 
collinear V-grooves as alignment canals for the fiber-based mirrors. The flat 
input mirror consists of a perpendicularly cleaved, dielectrically coated fiber 
(99.7\%$\pm0.05\%$). The concave cavity end mirror consists of a CO$_2$ laser 
ablated fiber tip with a high reflectivity dielectric coating 
(8ppm)\cite{BIBREF15}. The resulting cavity characteristics were measured to be: 
finesse $\mathcal{F}=1600$, optical quality factor 
$Q_\mathrm{opt}=3.5\times10^5$, and visibility $\gamma=96\%$. The optical output of the Fabry-P\'erot cavity in reflection is a function of the laser wavelength $\lambda$, yielding a signal $V(\lambda)$ at the photoreceiver given by, 
\begin{equation}
V(\lambda)=A\left(
1-\gamma\frac{1}{1+\left(\sin^2\left(\frac{\pi}{2\mathcal{F}}
\right)\right)^{-1}\,\sin^2\left(\frac{ 2\pi
L}{\lambda}\right)}\right)\,\,[
\mathrm{V}],
\label{Eqn4}
\end{equation}
where $A$ is the amplitude of the signal and $L$ is the optical cavity length. 

\begin{figure}[htp]
\centering
\includegraphics[width=0.38\textwidth, height=0.26\textwidth]{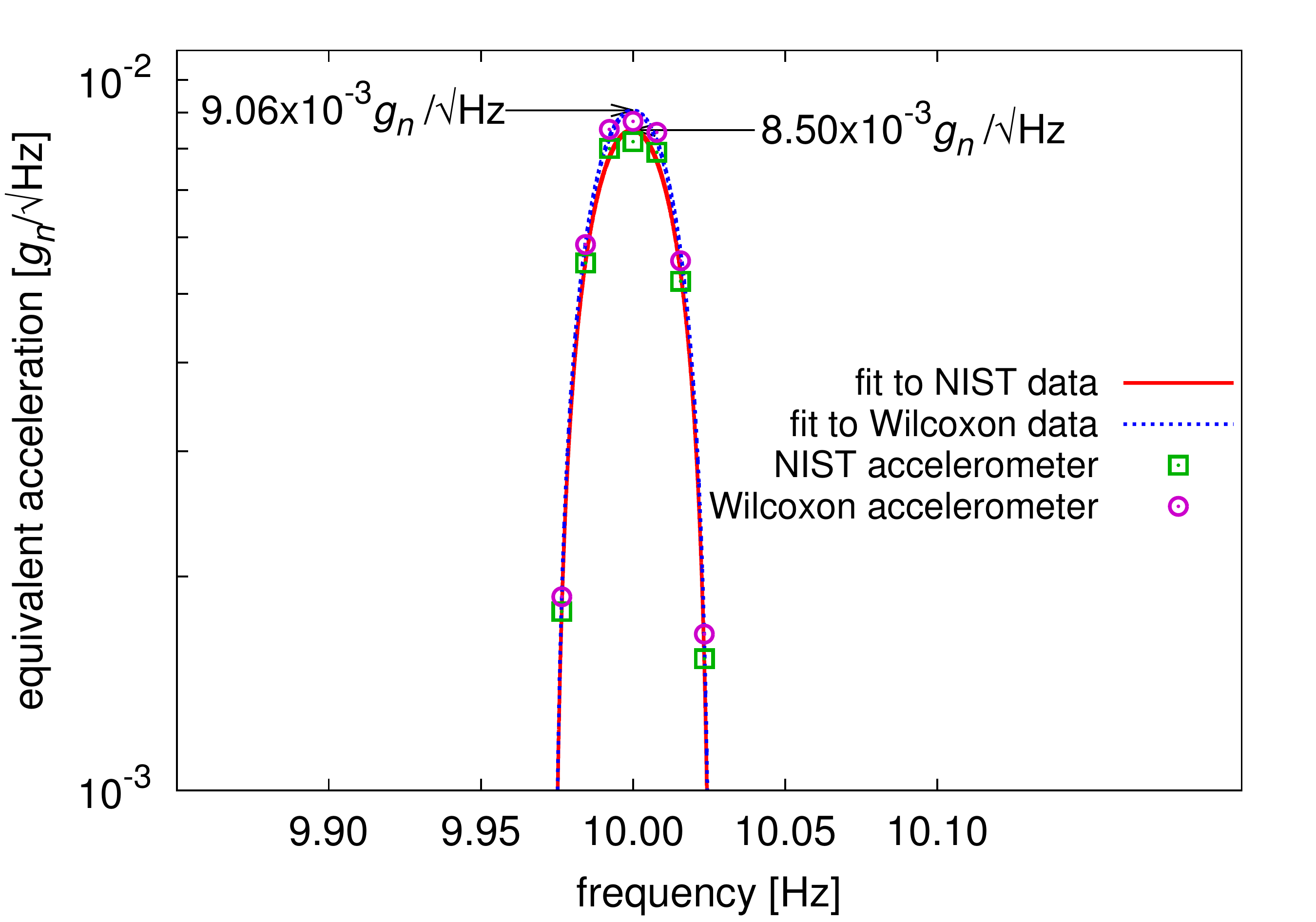}
\caption{\label{cal}
Measurement of an injected acceleration peak at 10\,Hz via a piezo-driven 
shaker to compare our device (NIST) with a calibrated commercial reference 
accelerometer Wilcoxon 731A/P31. The green (NIST) and magenta (Wilcoxon) dots 
show the data points recorded by the two devices. The red (NIST) and blue 
(Wilcoxon) traces show quadratic fits corresponding to each data set.}
\end{figure}

Several approaches can be implemented to conduct the optical length 
measurement. 
Typically, for field and long term operation systems it is necessary to actuate 
the laser frequency in closed loop in order to remain at the high sensitivity 
operation point around the cavity resonance\cite{BIBREF15a}. Heterodyne 
interferometry could also be implemented to measure the test-mass dynamics which 
would emerge as phase changes of the beat note that can be measured against a 
frequency reference\cite{BIBREF15b}. However, not all measurements require 
long-term stability and, particularly in our case, certain reference 
acceleration measurements, inter-device comparisons (as shown in 
Figure~\ref{cal}), and numerous calibrations can rely on the sufficient 
short-term stability provided by contemporary research laser systems, together 
with the SI-traceability capabilities offered by our reference acceleration 
measurement approach.

\begin{figure}[htp]
\centering
\includegraphics[width=0.38\textwidth, height=0.26\textwidth]{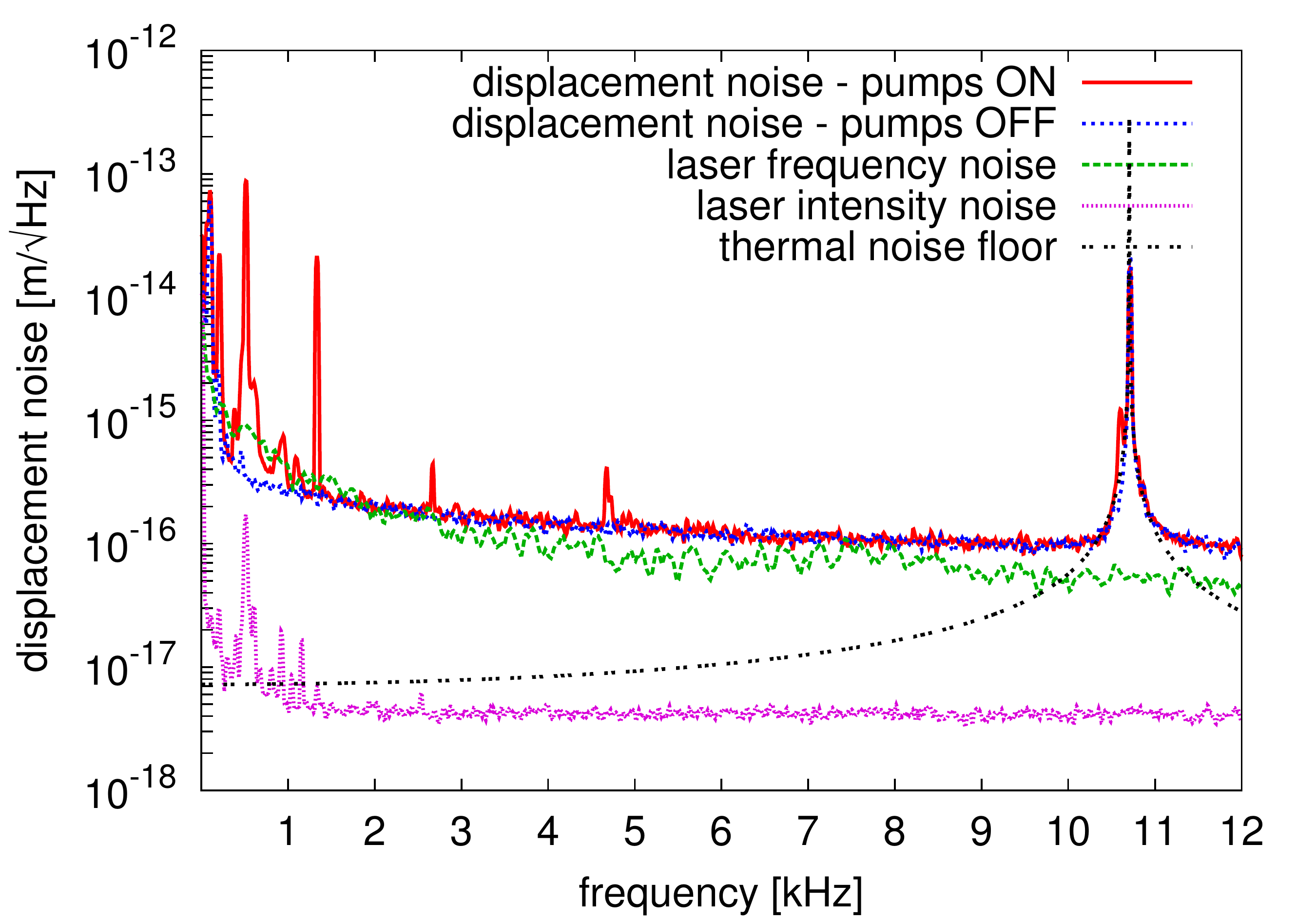}
\caption{\label{dispnoise}Linear spectral densities showing the sensitivity of the displacement measurements at a level of $200\,\mathrm{am}/\sqrt{\mathrm{Hz}}$. The red trace shows data measured while the vacuum pumps were running. The blue trace corresponds to the displacement sensitivity when the pumps have been shut down. The green and magenta traces are the equivalent displacement of the measured laser frequency and intensity noise, respectively. The black dotted trace is the computed thermal limit for the displacement measurement, considering the measured mechanical characteristics of the oscillator: $Q$, $m$, and $\omega_o$.} 
\end{figure}

Well-known modulation and AC detection schemes are 
typically used to avoid susceptibility to effects like laser power fluctuations 
which, however, have been measured to be negligible in our setup, as shown in 
Figure~\ref{dispnoise} (magenta trace--laser intensity noise). For these 
reasons, 
and striving for simplicity towards optomechanical reference acceleration 
standards, we have chosen a DC cavity readout to best accommodate the short 
length and moderate finesse properties of our optical sensor, for which a 
Pound-Drever-Hall\cite{BIBREF16} scheme is neither practical nor 
necessary. In this readout scheme, the highest sensitivity of the cavity can be 
achieved by tuning the laser wavelength to the points $\lambda_s$ where 
$\vert\mathrm{d}V(\lambda)/\mathrm{d}\lambda\vert$, reaches its maximum value. 
Given the wide tunability range of our laser 
(Agilent\,81600B:$\sim\,1450-1650$\,nm) the cavity length is directly measured 
by sweeping the laser wavelength over several nm and scanning the cavity 
resonances. The frequency separation between two adjacent resonance peaks is 
known as the Free Spectral Range (FSR) and its direct measurement yields the 
optical cavity length $L$ as,
\begin{equation}
L=\frac{c}{2\,\mathrm{FSR}},
\label{Eqn5}
\end{equation}
which amounts to 172\,$\upmu$m. Moreover, the remaining parameters of Equation~\ref{Eqn4},  the finesse $\mathcal{F}$, signal amplitude $A$ and visibility $\gamma$ can also be measured from this data. In itself, this FSR determination is a direct measurement of Equation~\ref{Eqn4}. The broadband displacement/acceleration measurement is conducted by measuring the laser intensity fluctuations returning from the cavity with a spectrum analyzer, whose master clock can also be referenced and traced to a frequency standard.  Cavity length fluctuations $\mathrm{d}L$ translate to laser wavelength fluctuations $\mathrm{d}\lambda$ in a resonant cavity as
\begin{equation}
\frac{\mathrm{d}L}{L}=\frac{\mathrm{d}\lambda}{\lambda}.
\label{Eqn6}
\end{equation}

The effect of cavity length fluctuations d$L$ on dynamic changes of the FSR are 
negligible given that $dL\ll~L$, being approximately 12 orders of magnitude 
smaller. By combining Equations~\ref{Eqn2},~\ref{Eqn4}~and~\ref{Eqn6}, it yields 
that cavity length changes driven by an acceleration $a_n$ can be measured via 
voltage fluctuations $V_n$ at the photoreceiver output as,
\begin{equation}
V_n = a_n\,\bigg|\bigg|-\frac{1}{\omega_o^2-\omega^2+i\,\frac{\omega_o}{Q}
\omega}\bigg|\bigg|\,\frac{\lambda_s}{L}\,
{\frac{\mathrm{d}V(\lambda)}{\mathrm{d}\lambda}}\bigg|_{\lambda_s},
\label{Eqn8}
\end{equation}
with
\begin{equation}
a_n = \sqrt{a_{th}^2+a_M^2+a_{\mathrm{EN}}^2},
\label{Eqn9}
\end{equation}
where the operating laser wavelength for maximum sensitivity $\lambda_s$ is an
SI-traceable quantity, and the function $\vert\mathrm{d}V(\lambda)/\mathrm{d}\lambda\vert$ can be either simultaneously measured with the FSR measurement, or analytically obtained upon measurement of the required parameters: $A$, $\gamma$, $\mathcal{F}$, $L$, and $\lambda_s$. The total measured acceleration $a_n$ consists of the 
following terms: a) the Brownian limited acceleration $a_{th}$ (see 
Equation~\ref{Eqn1}), b) equivalent acceleration $a_M$ of real test-mass 
dynamics, and c) apparent (absent of dynamics) acceleration $a_{EN}$ arising 
from excess noise in the detection system. Equation~\ref{Eqn8} demonstrates a 
straightforward link between the laser wavelength and acceleration-induced 
voltage fluctuations measured at the photoreceiver. All parameters 
($Q$,\,$\omega_o$,\,$\lambda_s$,\,$\mathrm{d}V(\lambda)/\mathrm{d}\lambda\vert_{
\lambda_s}$,\,$L$) are directly measured and are stable. This inherent relation 
between acceleration, the laser wavelength, and its explicit SI-traceability 
outlines the capabilities of absolute reference acceleration measurements of our 
device.

\begin{figure}[htp]
\centering
\includegraphics[width=0.35\textwidth, height=0.26\textwidth]{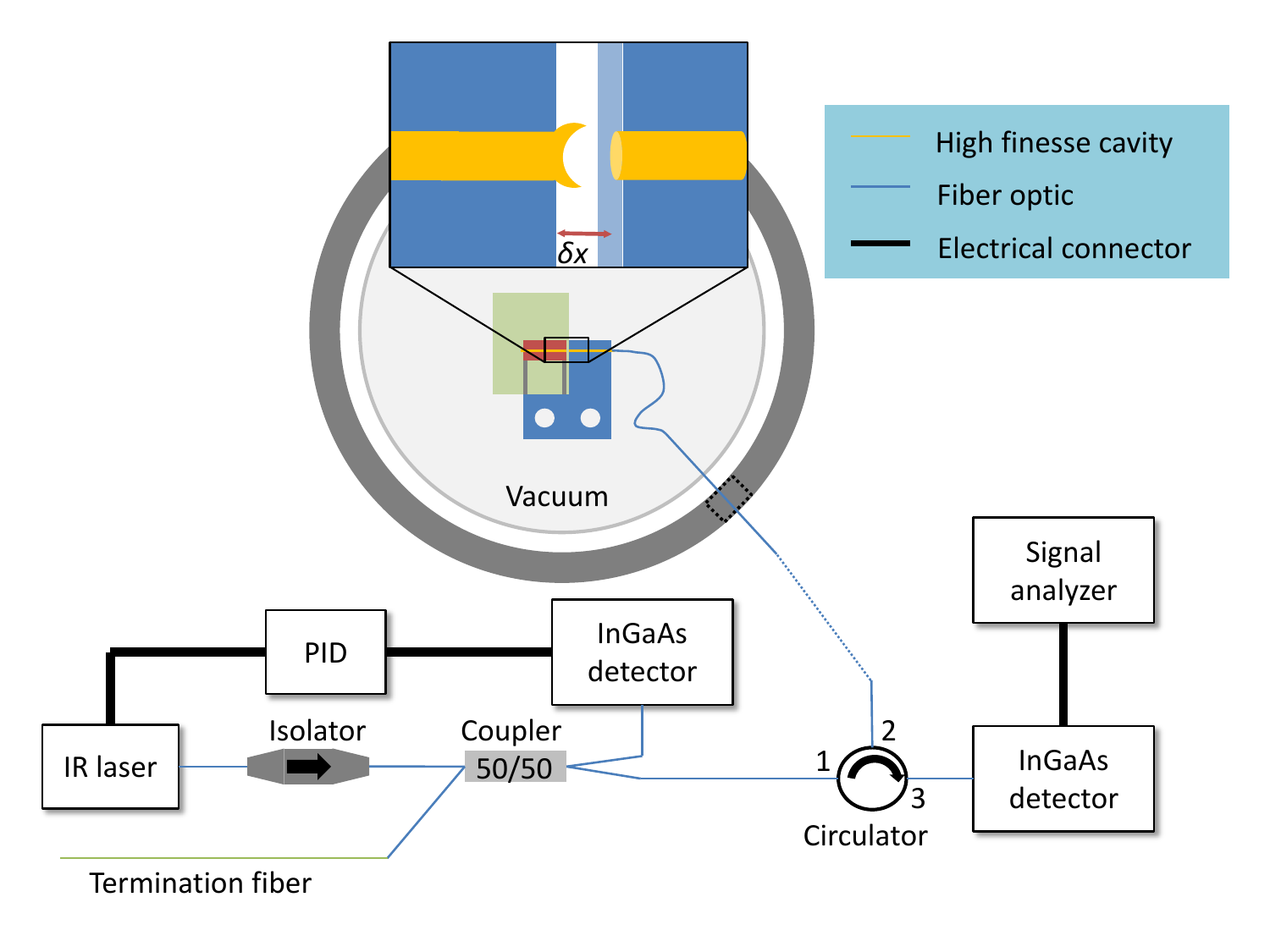}
\caption{\label{setup}Experiment setup schematics: the IR-laser is injected into the micro-cavity on the mechanical oscillator in vacuum through a fiber-isolator, a 50/50 fiber coupler, and a fiber circulator spliced to a fiber vacuum feed-through. One output of the 50/50 fiber coupler is connected to a intensity monitor photoreceiver that can be used for laser intensity stabilization. The output of the circulator launches the reflected cavity light onto the detection photoreceiver.}
\end{figure}

Our laboratory is located three floors 
below ground and the substrate was placed on a passive vibration 
isolation platform inside a vacuum chamber. This provides good mechanical 
isolation for acceleration noise measurements. The fiber optic cables were fed 
into the vacuum chamber through deformable capillaries that are clamped 
circumferentially around the cables for an air-tight seal. We used fusion 
splices to connect the input fiber to the fiber interferometer and laser sources 
and create a continuous strand of cable. A schematic of the test setup is shown 
in Figure~\ref{setup}.

Sensitivity measurements of displacement are shown in Figure~\ref{dispnoise}, reaching a resolution of $2\times10^{-16}\,\mathrm{m}/\sqrt{\mathrm{Hz}}$.

\begin{figure}[htp]
\centering
\includegraphics[width=0.38\textwidth, height=0.26\textwidth]{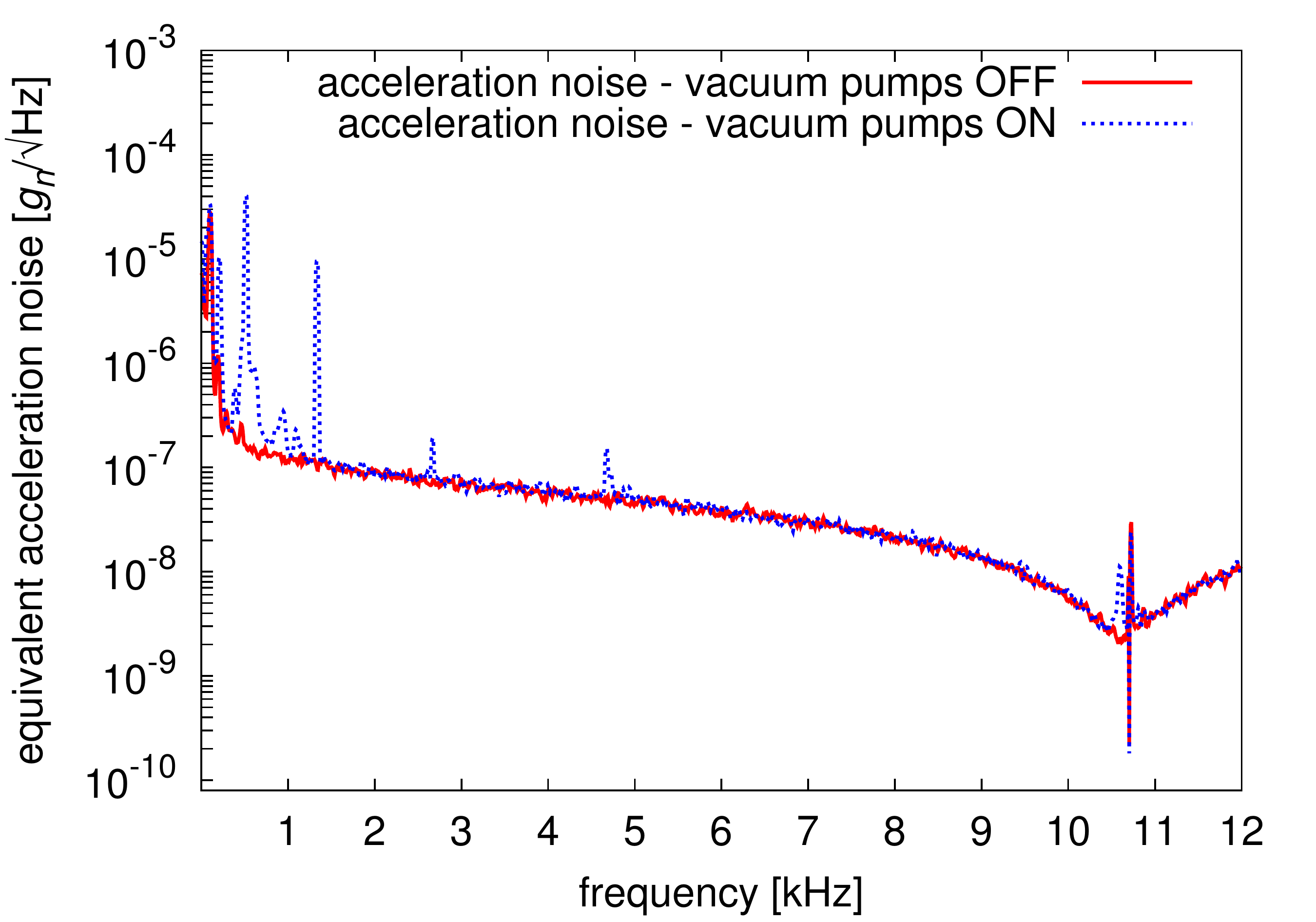}
\caption{\label{accnoise}Noise equivalent acceleration, demonstrating
sensitivities below $100\,\mathrm{n}g_n/\sqrt{\mathrm{Hz}}$ over 10\,kHz 
(1.5-12\,kHz).}
\end{figure}

The peaks shown in the red trace originate from the running vacuum pumps, and vanish when these are turned off while still maintaining a comparable vacuum level below 
$10^{-5}\,$mbar, as shown by the blue trace. In addition, we independently measured the laser intensity (magenta trace) and frequency (green trace) noise in order to identify 
the limiting noise sources of our system. The latter was measured via a 
mismatched armlength Mach-Zehnder fiber interferometer, using two 50/50 fiber couplers, and a 1\,km fiber spool in one arm. This fiber interferometer was assembled for the sole purpose of measuring the laser frequency noise and was operated outside the vacuum chamber, making it more susceptible to temperature changes and disturbances in the fibers. These environmental effects can account for the slightly higher noise below 2\,kHz. Given the rather high finesse accomplished on this type of cavity, we were able to suppress the laser intensity noise to a non-limiting level. Laser frequency noise, on the other hand, is clearly setting the noise 
floor. Figure~\ref{accnoise} shows the linear spectral density of the equivalent 
acceleration measurement. At lower frequencies of 10-100\,Hz we reach a 
comparable resolution to conventional devices at $\upmu g_n/\sqrt{\mathrm{Hz}}$ 
levels, and improving to sub-$\upmu g_n/\sqrt{\mathrm{Hz}}$ through 1\,kHz. 
Unprecedented sensitivities below 
$100\,\mathrm{n}g_n/\sqrt{\mathrm{Hz}}$ over 10\,kHz is achieved above 
1.5\,kHz, and better than $10\,\mathrm{n}g_n/\sqrt{\mathrm{Hz}}$ slightly 
above 9\,kHz over approximately 2\,kHz.

We are also able to exploit the extraordinary sensitivity 
enhancement provided by the mechanical resonance due to our capability to 
conduct a complete and accurate system identification over the entire 
observation bandwidth, allowing us to reach the Brownian acceleration limit 
around $3\,\mathrm{n}g_n/\sqrt{\mathrm{Hz}}$ level within 1\,kHz about the 
mechanical resonance frequency (10.7\,kHz).

The laser frequency noise sets the noise floor throughout the entire 
bandwidth. A 10\,kHz broadband detection sensitivity at the Brownian limit of 
$3\,\mathrm{n}g_n/\sqrt{\mathrm{Hz}}$ requires a laser frequency stability 
better than $10\,\mathrm{Hz}/\sqrt{\mathrm{Hz}}$ throughout the observation 
bandwidth. Laser frequency control systems capable of reaching 
these stability levels, have been previously demonstrated at the thermal 
limit\cite{BIBREF17}, but require complex dedicated laboratory 
instrumentation. In addition, ongoing 
research in laser physics\cite{BIBREF18,BIBREF19} outline the path for 
promising laser technologies of similar frequency stability that will be 
available in the near future which, combined with our optomechanical 
accelerometer, would enable broadband thermally-limited acceleration sensing.

Comparison measurements to a commercial reference accelerometer were conducted as a demonstration test of the acceleration sensing characteristics of our device. We have compared it against a calibrated commercial reference accelerometer, Wilcoxon 731A/P31\cite{BIBREF20}, which has a 3\,dB-bandwidth of 330\,Hz. We utilized a small piezo-driven shaker to inject a controlled acceleration to both devices at 10\,Hz. We chose this frequency because, from the only two frequency points (10\,Hz~and~327\,Hz) quoted in the calibration data sheet of the Wilcoxon device, this selected point reports 
unitary gain. The acceleration measurement was performed on one device at a time, mounted on the shaker. Figure~\ref{cal} shows the corresponding acceleration measurements of the injected signal.

The acceleration peaks measured by both devices come to excellent agreement. 
The slight discrepancy in amplitude of approximately 6\% can be attributed to 
large dimensional differences between the devices, which are of consideration in the mounting to the shaker table, and the effective acceleration field sensed by each test-mass. The Wilcoxon device has a mass of 670\,g and cylindrical dimensions of 6.2\,cm in diameter and 5.3\,cm in height. Our device has a sensing test-mass of 25\,mg, and a total mass that is 1000 times lower (approximately 600\,mg) than the Wilcoxon device, and it provides a ten times larger measurement bandwidth.   

In conclusion, we have demonstrated a simple and traceable optomechanical reference accelerometer to sensitivities better than $100\,\mathrm{n}g_n/\sqrt{\mathrm{Hz}}$ over an observation bandwidth of 10\,kHz. Our measurements of mechanical quality factor and mechanical resonance frequency demonstrate an $mQ$ product of 1\,kg and a Brownian acceleration limit of $3\,\mathrm{n}g_n/\sqrt{\mathrm{Hz}}$. We have also developed and incorporated an optical sensor based on a Fabry-P\'erot fiber micro-mirror cavity of finesse 1600 that reaches displacement sensitivities of $200\,\mathrm{am}/\sqrt{\mathrm{Hz}}$.

Many applications do not require acceleration sensing over such large bandwidths. Design trade-offs to the device concept presented here can be made to better meet the needs of sensitivity and observation bandwidth for other applications. The simplicity, compactness and traceable high-sensitivity performance over a wide bandwidth accompanied by built-in 
advanced laser-interferometric detection highlight our accelerometer for a wide variety of applications in optomechanics and quantum-limited light-matter interactions, absolute and reference accelerometry, including seismology, and ground and space-based gravimetry.\\

We thank J. Harris, N. Flowers-Jacobs, and S. Hoch for facilitating us samples of high reflective laser ablated fiber mirrors. We also thank C. Caves, N. Malalvala, K. Lehnert, G. Shaw and E. Polzik for helpful 
discussions. This work was supported by the NSF Physics Frontier Center at the 
JQI and DARPA QuASaR and the ARO under W911NF-11-1-0212.


%

\begin{thebibliography}{1}
\bibitem{BIBREF20} This paper identifies equipment for completeness. This is not an endorsement by the National Institute of Standards and Technology.
\end{thebibliography}
\end{document}